\documentclass[aps,pra,11pt]{revtex4-1}

\usepackage{graphicx}% Include figure files
\usepackage{dcolumn}% Align table columns on decimal point
\usepackage{bm}% bold math
\usepackage{color}
\begin{document}

%\preprint{AIP/123-QED}

\title{Inferring directed climatic interactions with renormalized partial directed coherence and directed partial correlation}% Force line breaks with \\

\author{Giulio Tirabassi}
 \altaffiliation[Now at ]{Guy Carpenter, Dublin, Ireland.}%Lines break automatically or can be forced with \\
\affiliation{ 
Universitat Polit\`ecnica de Catalunya, Departament de Fisica, Colom 11, 08222 Terrassa, Barcelona, Spain%\\This line break forced with \textbackslash\textbackslash
}%

\author{Linda Sommerlade}
 \email{l.sommerlade@abdn.ac.uk}
\affiliation{%
University of Aberdeen, Institute for Complex Systems and Mathematical Biology, Meston Walk, Aberdeen AB24 3UE, UK
%\\This line break forced% with \\
}%

\author{Cristina Masoller$^1$}
 \email{cristina.masoller@upc.edu}
% \homepage{http://www.fisica.edu.uy/~cris/}
\affiliation{%
Universitat Polit\`ecnica de Catalunya, Departament de Fisica, Colom 11, 08222 Terrassa, Barcelona, Spain%\\This line break forced% with \\
}%

\date{\today}% It is always \today, today,
             %  but any date may be explicitly specified

\begin{abstract}
Inferring interactions between processes promises deeper insight into mechanisms underlying network phenomena. Renormalised partial directed coherence (rPDC) is a frequency-domain representation of the concept of Granger causality while directed partial correlation (DPC) is an alternative approach for quantifying Granger causality in the time domain. Both methodologies have been successfully applied to neurophysiological signals for detecting directed relationships. This paper introduces their application to climatological time series. We first discuss the application to ENSO -- Monsoon interaction, and then apply the methodologies to the more challenging air-sea interaction in the South Atlantic Convergence Zone (SACZ). While in the first case the results obtained are fully consistent with present knowledge in climate modeling, in the second case the results are, as expected, less clear, and to fully elucidate the SACZ air-sea interaction, further investigations on the specificity and sensitivity of these methodologies are needed.
\end{abstract}

%  	Time series analysis, Inference methods, Complex systems,  	Properties and dynamics of the atmosphere; meteorology 
%Use showkeys class option if keyword
                              %display desired
\maketitle

\begin{quotation}
Many real-world complex systems can be represented in terms of complex networks of interacting elements. Typical examples include the brain, our climate, ecological systems, etc. Inferring the underlying interactions from observed data can yield new insight into network phenomena. Here we introduce the application of causal analysis tools to climatological time series. We demonstrate that these methods for the inference of causal interactions provide meaningfully information about long-range climate tele-connections.
\end{quotation}

\section{\label{sec:intro}Introduction}
Many real-world complex systems can be represented in terms of networks of interacting elements. Typical examples include the brain, our climate, ecological systems, etc. \cite{nets1,nets2,nets3}. In many situations the identification of causal interactions between elements is a challenging task, particularly in the case of weakly connected elements \cite{coupled_osc,ecosystems,palus_2013}. Therefore, a lot of research has been devoted to the development of reliable causality inference measures.

Partial directed coherence (PDC) \cite{pdc_1999,pdc_2001} is a frequency-domain representation of the concept of Granger causality \cite{granger_1969}. More recently, renormalised partial directed coherence (rPDC) was introduced to allow interpretation of the strength of connections\cite{ieee,pdc_2009}. In combination with state space modelling\cite{Sommerlade_2015} rPDC can cope with noisy data. Directed partial correlation (DPC) \cite{dpc_2003} is an alternative approach for quantifying Granger-causality in the time domain. In a similar way it can be combined with state space modelling to account for noise in the data. Both methodologies, (r)PDC and DPC, have been successfully applied to neurophysiological signals for detecting directed relationships \cite{pdc_2006,pdc_2009,pdc_linda,ieee,Saur_2010,den_Ouden_2012,par_corr_2016}. To the best of our knowledge, their potential for investigating climatological time series has not yet been explored. Because (r)PDC and DPC are robust against observational noise and have yielded meaningful insight in complex real data, we are motivated here to perform a first exploratory analysis to see if they can be useful also for climate data analysis. We first apply these methods to autoregressive (AR) processes, which have been commonly used in the literature to model climate data \cite{Hasselmann}. Then, we demonstrate the application of rPDC and DPC to the case of the El Ni\~no--Monsoon interaction. We conclude with an analysis of the more challenging air--sea interaction in the South Atlantic Convergence Zone (SACZ). For comparison we apply as a third diagnostic tool the Granger Causality Estimator (GCE)\cite{gce}, which has already been used on climatological time series\cite{Tirabassi_2015}.

Recently, several studies have brought insight into the Pacific-Indian Ocean interaction relevant for monsoonal dynamics \cite{grl_2011,runge_nat_comm_2015}. The El Ni\~no--Southern Oscillation (ENSO) is an atmospheric-oceanic phenomenon occurring mainly in the eastern equatorial Pacific Ocean. It can influence the climate also in a much greater part of the globe. With a periodicity of a few years, the easterly winds blowing along the equator in the Pacific Ocean become weaker, slowing down the rise of abyssal cold water in the front of Peruvian coasts. The net effect is a warming of the superficial waters of the eastern equatorial Pacific, that takes the name of El Ni\~no. The opposite phenomenon is known as La Ni\~na; in fact, Ni\~no and Ni\~na are opposite phases of ENSO.  While their frequency can be quite irregular, El Ni\~no and La Ni\~na events occur on average every two to seven years. Typically, El Ni\~no occurs more frequently than La Ni\~na \cite{noaa}.

The effects on the atmosphere of this huge quantity of heat accumulated on the ocean are various. In particular they contribute to an eastward shift of the convective patterns that affect the Maritime Continent, Australia and South America. This shift influences the so called \textit{Walker circulation}, a large cell-like circulation pattern placed across the Indian Ocean and the equatorial western and eastern Pacific Ocean \cite{book}. This circulation in turn affects the superficial temperature of the Indian Ocean, which impacts on the Indian monsoonal activity. 

In the context of climate networks, the interrelation between ENSO and Indian Summer Monsoon (ISM) activity has been represented by long-range links (tele-connections) between the Central Pacific and the Indian Ocean \cite{Kurths_2009,Barreiro_2011}. Recent work using a directionality index based on conditional mutual information has shown that the \textit{net} direction of interaction is from the Pacific to the Indian Ocean \cite{Deza_2015}, as expected from well-known climatic tropical variability patterns \cite{book,book2}.

In fact, one of the main climatic phenomena that is responsible for the Indian monsoon is the steep land-see temperature gradient. It triggers an inland stream of moisture that, interacting with the Indian orography, causes extreme rainfalls. Thus, the ENSO phenomenon affects the amount of rain falling over India, even if in an indirect way.  Conversely, changes in the superficial temperature of the eastern Indian Ocean, can also affect the Walker circulation. They impact on the easterly winds which by their variation can be partially responsible for the appearance of El Ni\~no. Therefore, it is important to investigate the existence of a bidirectional relationship between ENSO and ISM.

However, we can expect the forcing from ENSO to ISM to be significantly stronger, due to the difference in inertia and energy content of the two systems. Our hypothesis is the existence of mutual interactions that can be captured by the rPDC and DPC methods applied to relevant time-series. To test this, we consider two time-series that serve as proxy for ENSO and ISM activity: the NINO3.4 index and the All India Rainfall index, respectively. To apply the diagnostic tools, rPDC, DPC, and GCE, we assume that dynamics of the systems can be described by autoregressive processes. We show that, while the three measures give consistent results, DPC allows for a better discrimination of the strength of the weaker Indian monsoon $\rightarrow$ El Ni\~no interaction.

As a second, more challenging application we consider air--sea interactions in the south Atlantic convergence zone (SACZ). The SACZ is a large-scale convection phenomenon that takes place in austral summers across southern Brazil and the Atlantic Ocean. It can be thought of as a branch of the Inter-Tropical Convergence Zone (ITCZ). By carrying heavy rains, it can seriously affect the life of millions of people who live in the affected zones, as well as their economic activities. Due to its peculiar position, across subtropical and tropical latitudes, the nature of the SACZ phenomenon as a purely atmospheric phenomenon or as a coupled air-ocean one is still not fully understood.

Studies have shown that the subtropical South Atlantic Ocean may influence the evolution of the SACZ \cite{27, 30}. Even though the SACZ region is dominated by internal atmospheric variability, sea surface temperature anomalies can force a dipole of precipitation anomalies located mainly over the oceanic portion of the SACZ. Subsequent studies \cite{31,32} have suggested that the air-sea interaction is such that an initially stronger SACZ – due to internal atmospheric variability– induces an oceanic cooling that in turn negatively affects the convective precipitation, resulting in a negative feedback loop. However, air-sea interaction has been difficult to quantify both in observations and in model simulations and to date it is unclear how the circulation associated with the SACZ is influenced by surface ocean conditions. Disentangling air-sea interaction in the subtropics is a challenging task, and up to now to the best of our knowledge, no method has allowed a robust identification of this interaction in observational data. A first attempt to tackle this problem was recently presented by Tirabassi et al. \cite{Tirabassi_2015}. It was shown that the SACZ may exhibit a mixed behavior, that is, it can present at times as an ocean forced phenomenon, at others like a purely atmospheric, and at others as an entangled one. This was inferred from the analysis of a Granger causality estimator (GCE)\cite{gce} applied to the principal components (PCs) of the daily 500 mbar vertical velocity field in pressure coordinates anomalies (used as a proxy for precipitation and referred to as $w$) and the sea surface temperature anomalies (SST).  Here we will show that DPC reinforces the evidence that, in some years, SACZ may be forced by the SST.

This paper is organized as follows: Sec.~\ref{sec:metho} presents a simple example (two autoregressive processes of order 2, coupled in a unidirectional way) to illustrate the application of the three diagnostic tools: renormalised partial directed coherence (rPDC), directed partial correlation (DPC), and Granger causality estimator (GCE); Sec.~\ref{sec:ENSO} presents the application to the Indian monsoon and ENSO interaction and  Sec.~\ref{sec:sacz} presents the application to the air--sea interaction in the SACZ region. Finally, Sec.~\ref{sec:con} summarizes the conclusions.

\section{\label{sec:metho}Methods}

As a first diagnostic tool we use renormalized partial directed coherence (rPDC) in combination with a state space model. The state space model
\begin{eqnarray}
\label{eqn:stateSpace}
x(t) &=& \sum_{r=1}^{p} {a}(r) x(t-r) + \epsilon(t)\\
y(t) &=& \mathbf{C}x(t) + \eta(t)
\end{eqnarray}
enables the fitting of autoregressive processes (AR processes) $x(t)$ in the presence of observational noise $\eta(t)$. 
The interaction of the processes is captured in the coefficient matrices $a(r)$. To estimate rPDC, these matrices are Fourier-transformed and normalised.
For details on computing rPDC for time series at different frequencies we refer to the publication by Sommerlade et al.~\cite{Sommerlade_2015}. Here we illustrate the application of this concept with a simple example: the detection of the coupling direction between two autoregressive (AR) processes. We have chosen an AR model because it is commonly used in the literature to model climate data \cite{Hasselmann}.

We consider two AR processes, $M$ and $S$, of order 2, coupled in a unidirectional way, in a master-slave relationship:
\begin{eqnarray}
x_M^t&=&a_1^M x_M^{t-1}+ a_2^M x_M^{t-2}+ \epsilon_M^t \\
x_S^t&=&a_1^S x_S^{t-1}+ a_2^S x_S^{t-2}+ b_1 x_M ^{t-1} + b_2 x_M^{t-2}+ \epsilon_S^t
\end{eqnarray}
with $\textbf{a}^M=\textbf{a}^S=(0.7,0.1)$, $\textbf{b} =(0.05,0.01)$ and the standard deviation of the Gaussian noises $\epsilon^M= \epsilon^S=0.1$.  The system was integrated for 2000 steps and the last 1000 were used for the analysis. 
Observational noise of standard deviation 0.1 was added to the analysed time series.

\begin{figure}
\includegraphics[width=\columnwidth]{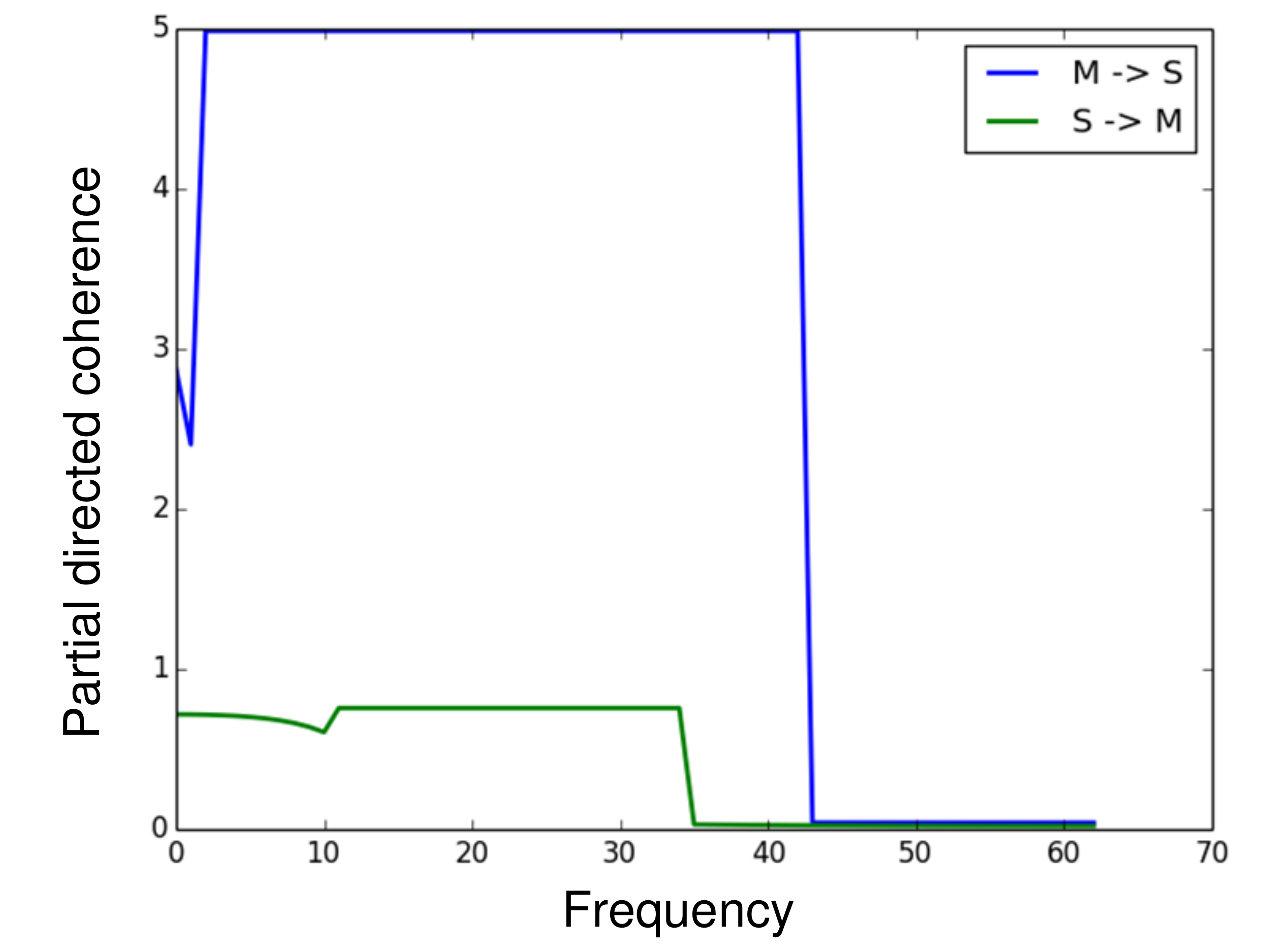}% chaos accepts pdf figs.
\caption{\label{fig:1} Renormalised partial directed coherence computed for two autoregressive processes, $M$ and $S$, unidirectionally coupled.}
\end{figure}

The results obtained with rPDC are depicted in Fig.~\ref{fig:1}. The rPDC spectrum in the case in which the master is forcing the slave ($M$ forcing $S$) is much higher than the one for the opposite direction of interaction.
This allows us to conclude, that $M$ drives $S$. However, since the processes under investigation show no clear oscillation frequency and rPDC is a frequency domain measure for Granger causality, interpretation is hampered.

The second diagnostic tool, directed partial correlation (DPC), allows us to analyze the problem in the time domain. We used the same state space model approach as for rPDC to fit an autoregressive process to the time series.
DPC then investigates the interactions directly in the time-domain by normalising the coefficient matrices $a(r)$ (Eq.~\ref{eqn:stateSpace}).
For computational details on the DPC method we refer to publications by Eichler et al.~\cite{dpc_2003}, Mader et al.~\cite{Mader_2008}, Saur et al.~\cite{Saur_2010} and den Ouden et al.~\cite{den_Ouden_2012}. The results of computing DPC with two lags, -1 and -2, are presented in Table~\ref{tab:table1}, where the evidence of a directed relation from $M$ to $S$ is strong, because DPC is significantly different from 0 only for M forcing S.

\begin{table}
\caption{\label{tab:table1}Directed partial correlation for two simulated AR processes, $M$ and $S$, with a master-slave relation.}
\begin{ruledtabular}
\begin{tabular}{c  cc  cc}
 &\multicolumn{2}{c }{$M$ forcing $S$}&\multicolumn{2}{c}{$S$ forcing $M$}\\
Lag&DPC&p-value&DPC&p-value\\
\hline
-1 & 0.09& 0.008 & 0.009 & 0.48\\
-2 & -0.03 & 0.008 & 0.004 & 0.48\\
\end{tabular}
\end{ruledtabular}
\end{table}

The third diagnostic tool is the Granger Causality Estimator (GCE) \cite{gce}. For this tool the AR processes are fitted to the data without a state space model. For computational details on the GCE method we refer to the publication by Tirabassi et al. \cite{Tirabassi_2015}. Computing the GCE for the two possible directions of interaction, we obtain the results presented in Table~\ref{tab:table3}. In this case, using a not particularly restrictive criterion for significance such as $p < 0.05$, we would conclude that $M$ is forcing $S$ and is also being forced by $S$. This result is interpreted as due to the presence of observational noise which was not accounted for when fitting the AR processes\cite{Sommerlade_2015, Newbold_1978, Nalatore_2007, Nolte_2008}. 

\begin{table}
\caption{\label{tab:table3} Granger Causality Estimator computed for the two autoregressive processes.}
\begin{ruledtabular}
\begin{tabular}{lcc}
Case& GCE &p-value\\
\hline
$M$ forcing $S$ & 0.016 & 0.001\\
$S$ forcing $M$ & 0.010 & 0.038\\
\end{tabular}
\end{ruledtabular}
\end{table}

A comparison of the three diagnostic tools allows us to conclude that, at least in this simple case, rPDC and DPC in combination with state space modelling better reconstruct the directed relationship between the two simulated AR processes than GCE based on AR fitting without a state space model.

\section{\label{sec:ENSO}ENSO and Summer Monsoon rainfall}

\begin{figure}
\includegraphics[width=\columnwidth]{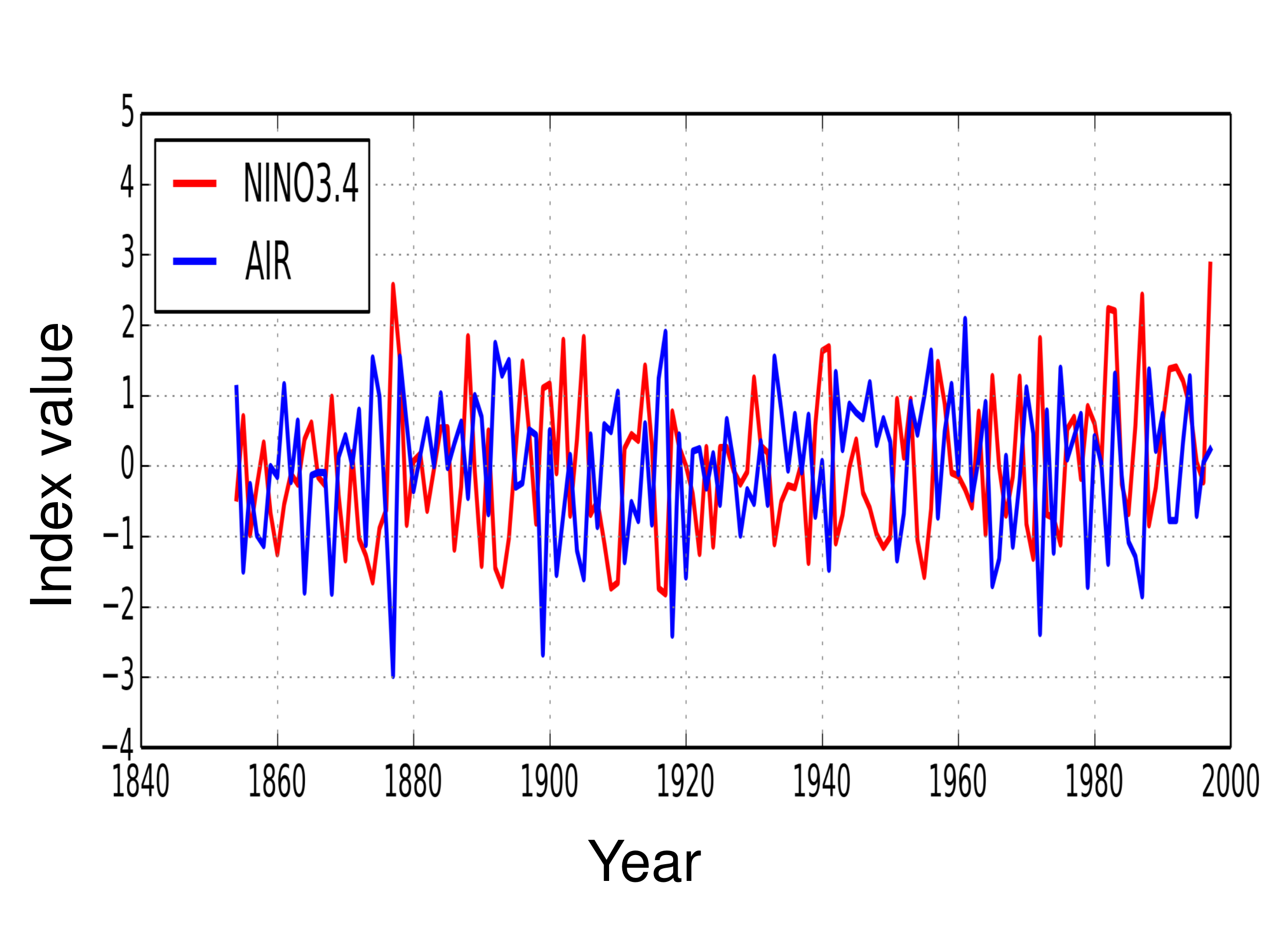}% chaos accepts pdf figs.
\caption{\label{fig:time_series} Time series of NINO3.4 index and All India Rainfall (AIR) index. Data from \textit{Climate Explorer}\cite{ce}.}
\end{figure}

Here we analyse the direction of interaction between two time-series that serve as proxy for ENSO and ISM activity. 
To represent ENSO dynamics, we consider the NINO3.4 index, which is the average sea surface temperature anomaly in the region from 5°N to 5°S and from 90°W to 150°W.
To represent the ISM activity, we have chosen the All India Rainfall (AIR) index. 
The data was taken from \textit{Climate Explorer}\cite{ce}: the NINO3.4 index begins in 1854, while AIR index begins in 1813. The shared length is 1728 months, from 1854 to 1998.  Both time series are shown in Fig. \ref{fig:time_series}. They are re-sampled to yearly resolution yielding 144 data points, which are normalized to zero mean and unit variance. 

From the physics of the problem it is not trivial to choose the order of the AR processes that are used to represent the data. Based on visual inspection of the fits, we heuristically selected a maximum lag of -4.
 
\begin{figure}
\includegraphics[width=\columnwidth]{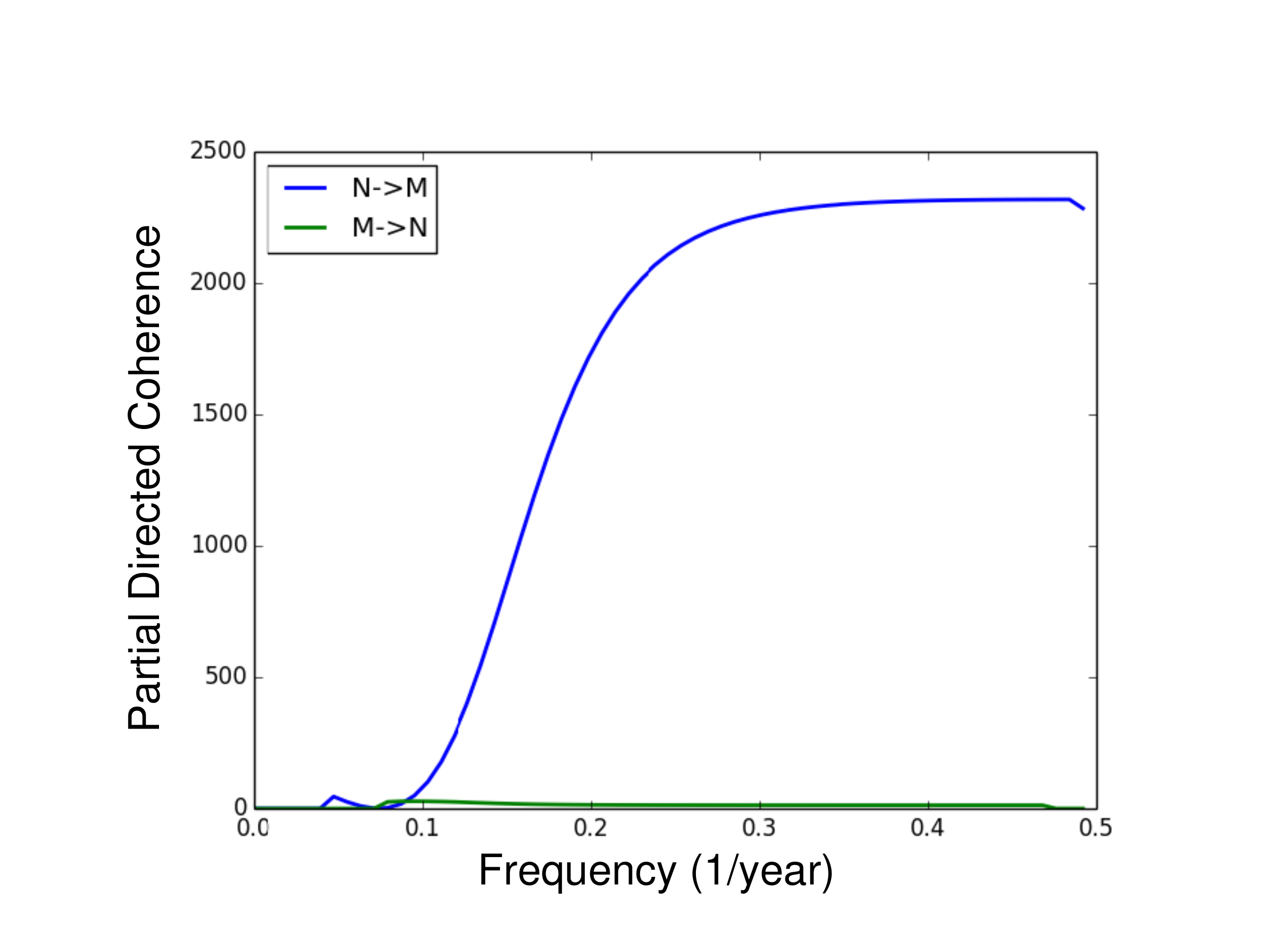}% chaos accepts pdf figs.
\caption{\label{fig:2} Renormalized partial directed coherence computed for the time series of the NINO3.4 index (ENSO, abbreviated by N) and the All India Rainfall index (ISM, abbreviated by M).}
\end{figure}

The results obtained with rPDC are displayed in Fig.~\ref{fig:2}. The connection from ENSO to ISM (N$\rightarrow$M) is much stronger than the connection in the opposite direction. However, interpretation of rPDC results is hampered because ENSO time-series has a characteristic oscillation frequency while the rainfall data does not.

The results for DPC are presented in Table~\ref{tab:table4}. High values in the hypothesis of ENSO forcing the ISM, and low values for the opposite direction are found. Moreover, all values are significantly different from 0.

\begin{table}
\caption{\label{tab:table4}Directed partial correlation for ENSO and ISM.}
\begin{ruledtabular}
\begin{tabular}{c cc cc}
 &\multicolumn{2}{c}{ENSO forcing ISM}&\multicolumn{2}{c}{ISM forcing ENSO}\\
Lag & DPC & p-value & DPC & p-value\\
\hline
-1 & 0.7& $<10^{-12}$ & -0.2 & $10^{-5}$\\
-2 & -0.9 & $<10^{-12}$ & 0.1 & $10^{-5}$\\
-3 & 0.7& $<10^{-12}$ & 0.1 & $10^{-5}$\\
-4 & -0.2 & $<10^{-12}$ & -0.1 & $10^{-5}$\\
\end{tabular}
\end{ruledtabular}
\end{table}

Finally, the results obtained with GCE, are presented in Table~\ref{tab:table6}. Both GCE values are significant. For the direction of ENSO forcing IMS, the p-value is much lower indicating that this connection is stronger than the opposite direction.

\begin{table}
\caption{\label{tab:table6} Granger Causality Estimator computed for the time series of the NINO3.4 index (ENSO) and the All India Rainfall index (ISM).}
\begin{ruledtabular}
\begin{tabular}{lcc}
Case& GCE &p-value\\
\hline
ENSO forcing IMS & 0.070 & $10^{-12}$\\
IMS forcing ENSO & 0.028 & $10^{-4}$\\
\end{tabular}
\end{ruledtabular}
\end{table}

All three methods agree in the sense that both directions are present and ENSO forcing IMS is the stronger connection. However, the results obtained with DPC additionally allow to interpret delays in the connectivity.

\section{\label{sec:sacz}The South Atlantic Convergence Zone} 

\begin{figure}
\includegraphics[width=\columnwidth]{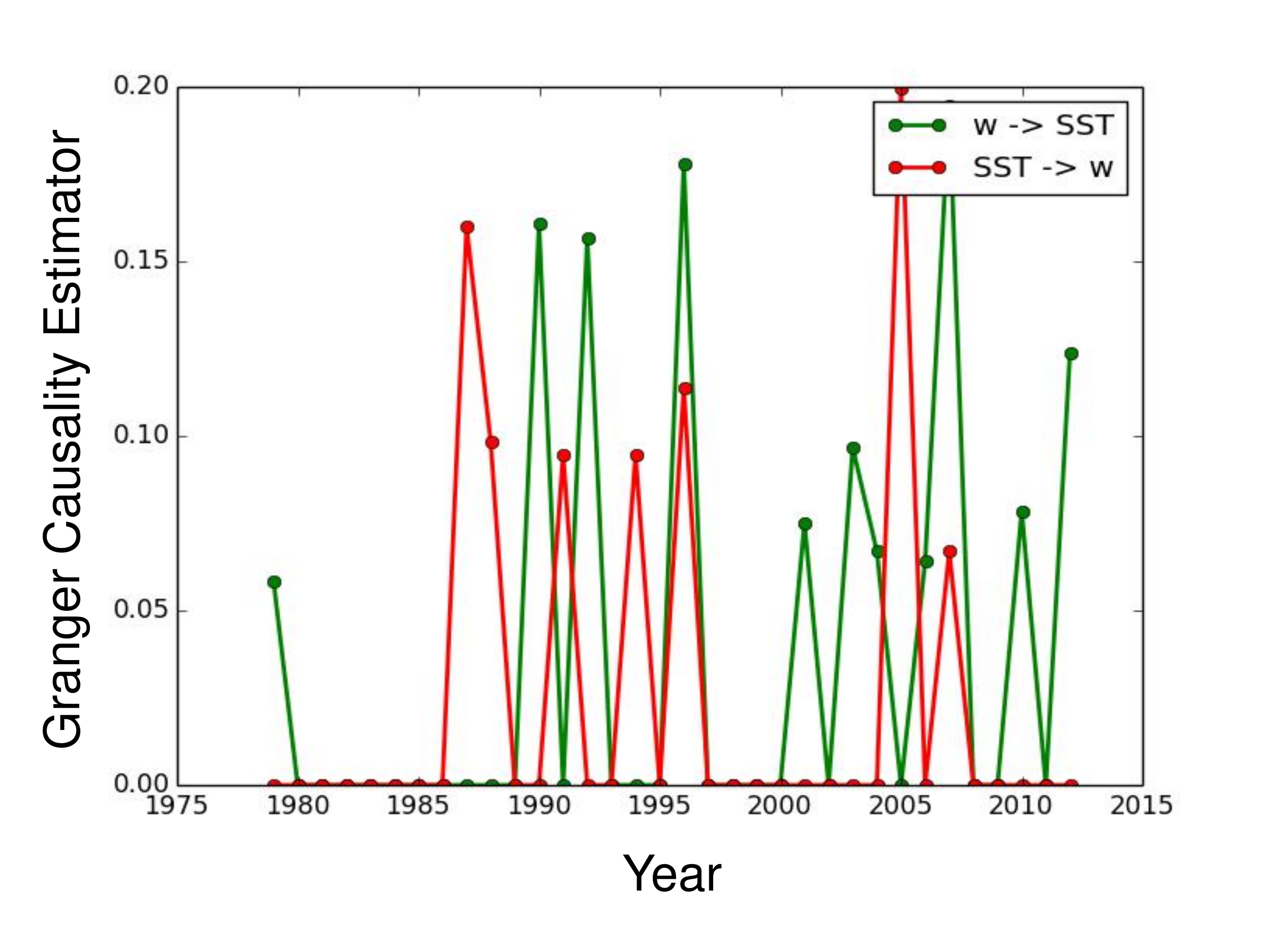}% chaos accepts pdf figs.
\caption{\label{fig:3} Granger causality estimator computed for the time series of the daily 500 mbar vertical velocity field in pressure coordinates anomalies (referred to as $w$) and the sea surface temperature anomalies (SST) in the south Atlantic convergence zone (SACZ).}
\end{figure}

As discussed in the Introduction, recent evidence suggests that air--sea interaction in the SACZ region can alternate, with years in which the forcing is mainly directed from the atmosphere to the ocean, years in which the ocean forces the atmosphere, years in which the interaction is mutual and years in which the interaction is not significant \cite{Tirabassi_2015}.
This was inferred from the analysis of the principal components (PCs) of the daily 500 mbar vertical velocity field in pressure coordinates anomalies (used as a proxy for precipitation and referred to as $w$) and the sea surface temperature anomalies (SST). 
Computing GCE between these two time series (see \cite{Tirabassi_2015} for details), we obtained the results presented in Fig.~\ref{fig:3}. 
For the sake of clarity, the non-significant values of GCE below the 0.1 p-value threshold have been set equal to 0. 
It can be seen, that it is possible to observe time intervals in which the sea is forcing the atmosphere, i.e.~years in which only the SST~$\rightarrow w$ values are significant (e.g. 1987), and time intervals in which the atmosphere is forcing the ocean, i.e.~years in which only the $w\rightarrow$~SST values are significant (e.g. 2001).
Moreover, there are years in which the systems are mutually coupled, i.e.~when both SST~$\rightarrow w$ and $w\rightarrow$~SST values are significant (e.g. 1996) and years in which they are uncoupled, i.e.~when none of the two values is significant (e.g. 1999). 

\begin{figure}
\includegraphics[width=\columnwidth]{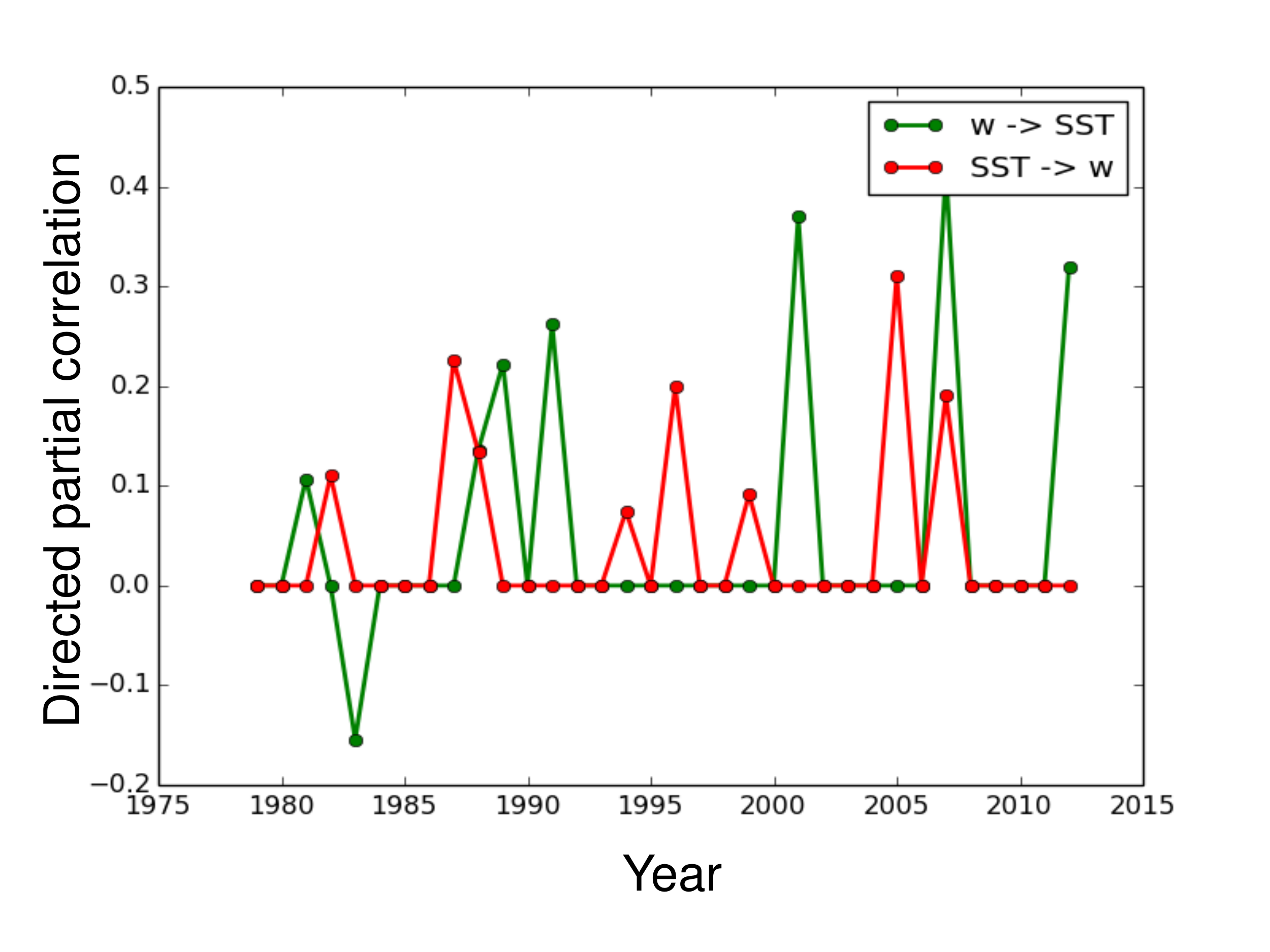}% chaos accepts pdf figs.
\caption{\label{fig:4} Directed partial correlation computed for the time series of 500 mbar vertical velocity field in pressure coordinates anomalies (referred to as $w$) and the sea surface temperature anomalies (SST) in the SACZ.}
\end{figure}

In order to verify these conclusions, we compute the directionality between the two time series with directed partial correlation. The decision of the optimal model order is not straightforward. We heuristically chose lag -1 based on visual inspection of the fits. A more detailed analysis of the choice for the model order is left for future work. 
The results obtained with DPC are displayed in Fig.~\ref{fig:4}.

\begin{figure}
\includegraphics[width=\columnwidth]{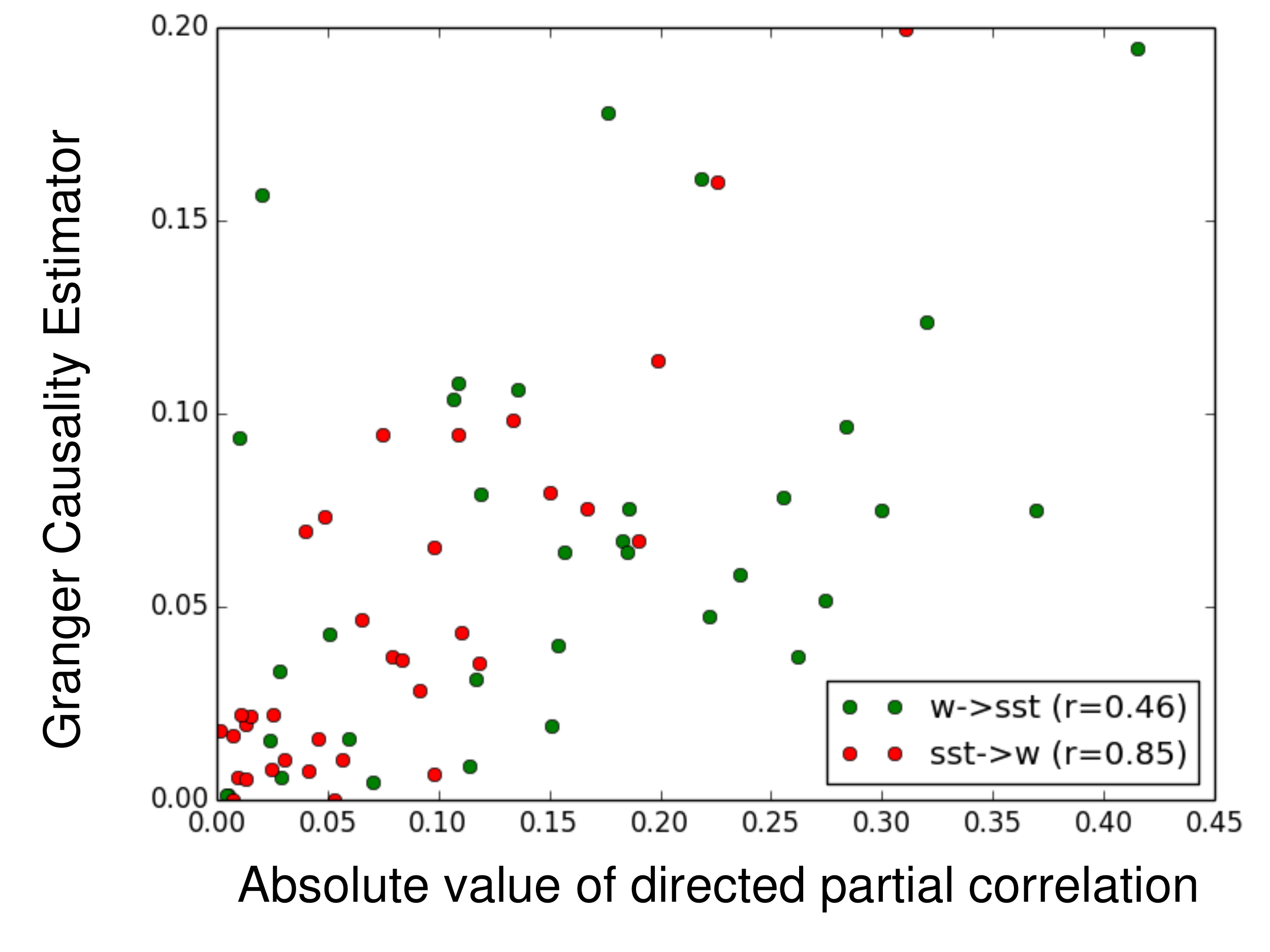}% chaos accepts pdf figs.
\caption{\label{fig:5} Granger causality estimator vs. absolute value of directed partial correlation for every year, based on the time series of 500 mbar vertical velocity field in pressure coordinates anomalies (referred to as $w$) and the sea surface temperature anomalies (SST) in the SACZ.}
\end{figure}

There are some similarities between the plots of GCE and DPC versus time, which become clearer in a scatter plot of GCE versus DPC for every year (Fig.~\ref{fig:5}). We observe in Fig.~\ref{fig:5} that the SST~$\rightarrow w$ case is strongly correlated between the two measures (r = 0.85), 
reinforcing the evidence that indeed the SACZ may be forced by the SST in some years. 
In contrast, the $w\rightarrow$~SST case shows a weaker correlation (r = 0.46) maybe due to the different effect of observational noise on the two diagnostic tools. For DPC we used a state space model approach that takes into account observational noise. Thus, false positive conclusions merely due to the presence of noise are eliminated.

No rPDC analysis was performed on SACZ because the time series showed no clear oscillation frequency for the processes under investigation. Without a well-defined oscillatory behavior the spectral method is not expected to yield meaningful insight.

\section{\label{sec:con}Conclusions}

We have used three diagnostic tools [renormalized partial directed coherence (rPDC), directed partial correlation (DPC), and Granger causality estimator (GCE)] to analyze the directionality of climatological interactions from time series of relevant climatological indices. As a first example we considered the ENSO--Monsoon mutual interaction. We found that the three measures give consistent results, which are also fully consistent with present knowledge in climate modeling.

As a second example we analyzed the air-sea interaction in the South Atlantic Convergence Zone (SACZ). A comparison of DPC and GCE reinforced the evidence that, in some years, SACZ may be forced by SST. Interestingly, there is one year (1983) in which the pressure coordinates anomalies ($w$) seems to negatively influence the sea surface temperature anomalies (SST). Such a negative effect cannot be detected by GCE as this measure does reveal if the connection is positive or negative. However, detailed investigations on the specificity and sensitivity of these methodologies are needed in order to obtain reliable insights on climatic phenomena.
The reported SACZ p-value is a single test significance level. While multiple tests are often performed, a correction for multiple testing in this case was not done because we are testing, for different years, how often (not if at all) there is a link. We are confident in the results because for random fluctuations we would expect about 7 spurious GCE values above the significance threshold, while we found 18. The results presented here are a first application and future work could focus on detailed interpretation as well as on using surrogate data and other means to validate the findings.

An important drawback of these methodologies is that the results depend on the order of the AR process (if it is too low or too high there is overfit or underfit of the empirical data). As stated before, as a first approach we have chosen the order of the AR process heuristically; however, in order to further advance the applicability of these methods to climate data, a detailed analysis of the influence of the order of the AR process is needed. While it is possible, in general, to interpret DPC for different lags, the interpretation is  straight forward only if the underlying process is indeed an AR process. Several factors such as non-linearities as well as observational noise (even if it is accounted for in the estimation procedure) can hamper the interpretation. A tailored simulation study would be needed to further investigate interactions at different lags, which is beyond the scope of this article.

The diagnostic tools used here are applicable to a broad class of systems, for inferring their network structures. As discussed in the Introduction, PDC has been successfully applied, for example, to fMRI data for the detection of directed information flow \cite{ieee}; and classical Granger causality has been applied to multivariate climate datasets (see \cite{palus} by Prof. Palus and co-workers in this special issue). The main challenge is a reliable parameter estimation, which becomes difficult if the network is large because many coefficients need to be estimated and typically only limited data are available.

\begin{acknowledgments}
This work was supported in part by Spanish MINECO/FEDER (FIS2015-66503-C3-2-P) and ITN LINC (FP7 289447). C. M. also acknowledges partial support from ICREA ACADEMIA.
\end{acknowledgments}

\nocite{*}
%\bibliography{aipsamp}% Produces the bibliography via BibTeX.

\end{document}